# PERANCANGAN BASISDATA SISTEM INFORMASI PENGGAJIAN
**(Studi Kasus pada Universitas 'XYZ')**

**Oleh: Leon Andretti Abdillah**
**Dosen Universitas Bina Darma, Palembang**

***Abstracts***: *The purpose of this research is to design database scheme of information system at 'XYZ' University. By using database design methods (conceptual scheme, logical scheme, & physical scheme) the writer designs payroll information system. The physical scheme is compatible with Borland Delphi Database Engine Scheme to support the implementation of the I.S. After 3 (three) steps we get 7 (seven) tables, dan 6 (six) forms. By using this shemce, the system can produce several reports quickly, accurately, efficiently, and effectively.*

## 1. PENDAHULUAN

Salah satu teknologi yang sedang berkembang dengan pesat saat ini adalah teknologi informasi/komputer. Kemajuan yang berlangsung cepat, dapat ditinjau baik dari segi perangkat keras (*hardware*), perangkat lunak (*software*), maupun perkembangan kualitas sumber daya manusianya (*brainware*). Hal ini dimungkinkan karena teknologi komputer mampu berkolaborasi dengan banyak bidang ilmu lainnya.

Perkembangan teknologi informasi dapat meningkatkan kinerja dan memungkinkan berbagai kegiatan dapat dilaksanakan dengan cepat, tepat dan akurat, sehingga akhirnya akan meningkatkan produktivitas. Perkembangan teknologi informasi memperlihatkan bermunculannya berbagai jenis kegiatan yang berbasis pada teknologi ini, seperti *egovernment, e- commerce, e-education, emedicine, e-e-laboratory*, dan lainnya, yang kesemuanya itu berbasiskan elektronika. (Wardiana, http://eprints.rclis.org/archive/).

Kelebihan yang luar biasa dari Teknologi Informasi, sudah semestinya dimanfaatkan oleh beragam pihak untuk mengelola beragam aktivitas institusinya dengan baik, terencana, dan terdokumentasi dengan maksimal.

Dalam menghadapi tantangan abad ke-21 atau milenium ketiga, yang ditandai dengan adanya era globalisasi, era reformasi dan era kompetisi, peranan Sumber Daya Manusia (SDM) untuk kemajuan organisasinya menjadi hal yang amat penting karena Sumber Daya Manusia memiliki peran yang sangat strategis dalam pencapaian tujuan organisasi di lingkungan perguruan tinggi yang pada saat ini dan masa datang dituntut menjadi organisasi yang otonom sehingga berdampak terhadap manajemen Sumber Daya Manusia. Manajemen SDM yang dibutuhkan adalah manajemen yang relevan dengan tuntutan lingkungan eksternal perguruan tinggi yang selalu berubah dan berkembang ke arah yang lebih kompetitif dan *profitable*. (Cornelia J. Benny, 2004).

Dunia pendidikan merupakan lembaga yang berperan utama untuk menghasilkan sumber daya manusia agar mampu menguasai ilmu pengetahuan dan teknologi. Universitas 'XYZ' Palembang merupakan salah satu lembaga pendidikan formal yang menghasilkan tenaga ahli dibidang ilmu komputer, ekonomi, teknik, bahasa dan sastra, psikologi, komunikasi, serta manajemen.

Seiring dengan bertambahnya usia lembaga Universitas 'XYZ' ini ternyata mampu meningkatkan jumlah mahasiswa yang menimba ilmu di kampus ini. Agar para mahasiswa tersebut mempunyai kualitas yang tinggi, harus didukung oleh tenaga pengajar yang handal, teknologi pengolah data yang canggih, serta manajemen pengelolaan sumber daya yang akan melayani kegiatan proses belajar mengajarnya. Saat ini tenaga pengajar di Universitas 'XYZ' terdiri atas sejumlah Profesor, Doktor (S-3), Master/Magister (S-2), dan Sarjana (S-1) yang berkompeten, baik lulusan dalam maupun luar negeri.

Untuk kelancaran proses penyelenggaraan pendidikan dibutuhkan tidak hanya tenaga pengajar yang berkualitas dan seimbang, tetapi juga termasuk tenaga-tenaga pendukung lainnya seperti: tenaga administrasi dan keuangan, laboran, teknisi, pustakawan, keamanan, dll. Sebagai aset yang sangat penting Sumber Daya Manusia yang ada perlu ditingkatkan pengelolaannya dengan lebih baik lagi, khususnya yang berhubungan dengan Sistem Penghargaan/Kompensasi atas kinerjanya.

Dari pengamatan yang dilakukan penulis pada bagian yang terkait dengan pengolahan data gaji, masalah yang dihadapi adalah: 1) masih belum akuratnya penghitungan gaji bersih para dosen, 2) kurang cepatnya informasi mengenai slip gaji dosen, dan 3) *up-date* data yang terjadi pada seorang dosen belum secara otomatis ter-integrasi pada bagian penggajian. Kesemuanya ini dikarenakan penggunaan *database* (basisdata) tidak diterapkan dengan baik, sehingga banyak kegiatan yang semestinya dapat dilakukan 1 (satu) kali saja, namun dilakukan secara berulang.

Penulis dapat merumuskan sebagai berikut: Bagaimana membangun skema perancangan basisdata (*datatabase*) Sistem Informasi Dosen pada Universitas 'XYZ' Palembang dengan menggunakan *Borland Delphi* 7.0?.

Adapun tujuan dari penelitian ini adalah membuat skema perancangan *database* (basisdata) untuk Sistem Informasi Penggajian pada Universitas 'XYZ'. Sedangkan Manfaat dari penelitian ini adalah: 1) Membantu bagian pengembangan perangkat lunak dalam membuat Sistem Informasi Penggajian, dan 2) Memudahkan bagian keuangan dalam mengolala data penggajian dan dengan mudah mendapat informasi yang dibutuhkan.

## 2. TINJAUAN PUSTAKA

Untuk memudahkan proses perancangan skema *database* Sistem Informasi Dosen pada Universitas 'XYZ', perlu diketahui sejumlah teori atau penelitian yang pernah dilakukan, baik oleh penulis sendiri maupun dari rujukan lain.

### Perancangan Basisdata (*Database Design*)

Proses perancangan *database* merupakan bagian dari *micro lifecycle*. Sedangkan kegiatan-kegiatan yang terdapat di dalam proses tersebut diantaranya : pengumpulan data dan analisis, perancangan *database* secara konsepsual, pemilihan DBMS, perancangan *database* secara logika (*data model mapping*), perancangan *database* secara fisik, dan implementasi sistem *database*. Sekarwati (2001) dalam Abdillah (2003:18).

Sedangkan kegiatan utama dalam perancangan suatu *database* adalah: 1) perancangan basisdata secara konsepsual (*conceptual scheme design*), 2) perancangan basis data secara logika (*logical design*), dan 3) perancangan basisdata secara fisik (*phisycal design*).

Tujuan perancangan basisdata : 1) untuk memenuhi informasi yang berisikan kebutuhan-kebutuhan user secara khusus dan aplikasi-aplikasinya, 2) memudahkan pengertian struktur informasi, dan 3) mendukung kebutuhan-kebutuhan pemrosesan dan beberapa obyek penampilan (*response time, processing time,* dan *storage space*). (Abdillah, 2003:20).

### Sistem Informasi

Disarikan dari Robert A. Leitch dan K. Roscoe Davis dalam Jogiyanto (2000:11) serta Eko Indrajit (2001:3), secara sederhana, Sistem Informasi merupakan kumpulan komponen yang saling berhubungan untuk mengolah *input* (data) menjadi *output* (informasi) sehingga dapat memenuhi kebutuhan pemakai. (Abdillah, 2006:11).

Komponen-komponen utama dalam suatu sistem informasi berbasiskan komputer terdiri dari: 1) *Database*, 2) *Database software*, 3) Aplikasi *software*, 4) *Hardware* komputer termasuk media penyimpanan, dan 5) Personal yang menggunakan dan mengembangkan system. (Abdillah, 2003:18).

**Penggajian**

Kompensasi/upah adalah imbalan atas jasa yang dapat berbentuk secara langsung (berbentuk uang), atau secara tidak langsung (misalnya asuransi kesehatan, fasilitas liburan). Dari penjelasan tersebut dapat disimpulkan bahwa dimensi dari kompensasi ada 2 (dua), yaitu : kompensasi secara langsung (*direct compansation*), dan kompensasi secara tidak langsung (*indirect compensation*). (Abdillah, 2006:33).

Gaji adalah suatu bentuk balas jasa ataupun penghargaan yang diberikan secara teratur kepada seorang pegawai atas jasa dan hasil kerjanya. Gaji sering juga disebut sebagai upah, dimana keduanya merupakan suatu bentuk kompensasi, yakni imbalan jasa yang diberikan secara teratur atas prestasi kerja yang diberikan kepada seorang pegawai. Perbedaan gaji dan upah hanya terletak pada kuatnya ikatan kerja dan jangka waktu penerimaannya. Seseorang menerima gaji apabila ikatan kerjanya kuat, sedang seseorang menerima upah apabila ikatannya kerjanya kurang kuat. Dilihat dari jangka waktu penerimaannya, gaji pada umumnya diberikan pada setiap akhir bulan, sedang upah diberikan pada setiap hari ataupun setiap minggu. Dalam hal ini, pengertian gaji untuk seterusnya disebut sebagai gaji pokok. (http://kuliah.dinus.ac.id/edi-nur/lembar01.html/lembar01.html).

**Dosen**

Dosen adalah tenaga pengajar di perguruan tinggi yang berdasarkan pendidikannya diangkat oleh penyelenggara perguruan tinggi. Dosen dapat dikategorikan menjadi dosen tetap dan dosen luar biasa. Dosen tetap ialah dosen yang diangkat dan ditempatkan sebagai tenaga tetap pada perguruan tinggi bersangkutan, sedangkan dosen luar biasa adalah dosen yang bukan tenga tetap (tidak tetap) selama jangka waktu tertentu. (Koordinator Kopertis II, 2005:1).

**Borland Delphi**

*Borland Delphi 7.0 (Delphi)*, adalah paket bahasa pemrograman yang bekerja dalam sistem operasi *Windows*. *Delphi* merupakan bahasa pemrograman yang mempunyai cakupan kemampuan yang luas dan sangat canggih. Berbagai aplikasi dapat dibuat dengan *Delphi*, termasuk aplikasi untuk pengolah teks, grafik, angka, *database* dan aplikasi *web*.

*Delphi* menyediakan fasilitas yang lengkap untuk mengelolah *database*. Berbagai format *database* dapat diolah dengan *Delphi*, misalnya *database* dengan format *Paradox, dBase, MS-Access, ODBC, SyBASE, Oracle* dan lain-lain.

Sebelum *Delphi*, *Borland* sudah lama mengeluarkan program untuk manajemen *database* yang sangat terkenal, yaitu program *Paradox*. Dengan *Delphi*, kemampuan yang ada pada program *Paradox* menjadi lebih baik dan makin sempurna. Dalam format *Paradox*, satu *file database* hanya berisi satu tabel *database*. Jadi berbeda dengan format *Microsoft Access*, yang memungkinkan membuat beberapa tabel dalam satu *file database*. Dalam penelitian kali ini penulis menggunakan *Paradox* sebagai manajemen *database*nya.

## 3. METODOLOGI PENELITIAN

**Lokasi Penelitian**

Lokasi penelitian dilakukan pada Universitas 'XYZ', beralamat di Jalan Jenderal Ahmad Yani Plaju, Palembang. Waktu yang dibutuhkan untuk mendapatkan data serta pengolahannya adalah selama 1 (satu) bulan.

**Metode Pengumpulan Data**

Untuk mendapatkan data dan informasi yang diperlukan, penulis menggunakan metode deskriptif, yaitu dengan cara mengumpulkan data dan informasi di lingkungan Universitas 'XYZ'. Penulis mengadakan penelitian dengan cara sebagai berikut: 1) Observasi, dengan melakukan pengamatan langsung pada bagian Administrasi dan Keuangan pada Universitas 'XYZ', 2) Wawancara, dilakukan langsung dengan a) Kasubbag. Keuangan, b) Staf Administrasi Penggajian, dan c) Bagian Administrasi Umum, dan 3) Studi Pustaka, dilakukan dengan membaca Buku Pedoman Dosen, Buku Manajemen Sumber Daya Manusia, Buku Basisdata, serta Buku Pedoman Penggunaan *Borland Delphi 7.0*.

**Metode Perancangan Basisdata (*Database Design*)**

Metode perancangan basisdata (*database design*) yang digunakan adalah: 1) perancangan basisdata secara konsepsual (*conceptual scheme design*), 2) perancangan basisdata secara logika (*logical design*), dan 3) perancangan basisdata secara fisik (*phisycal design*).

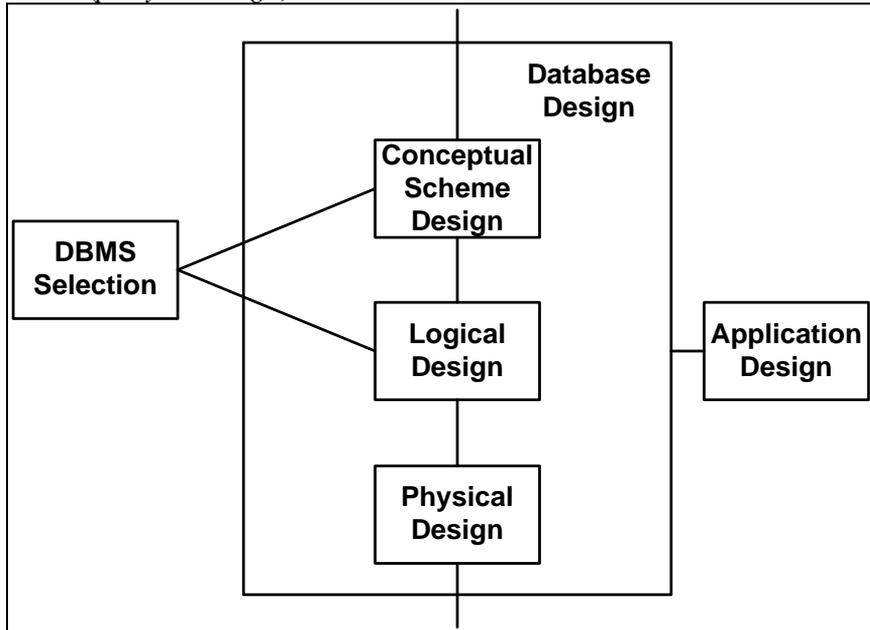

Disarikan dari Siklus Kehidupan Database sebagai Siklus Kehidupan Mikro.
Sekarwati (2001) dalam Abdillah (2003:19).

**Gambar 1. Skema Konsepsual (*Conceptual Scheme*)**

**Perancangan Basisdata secara Konsepsual (*Conceptual Design*)**

Perancangan skema konsepsual: menguji kebutuhan-kebutuhan data dari suatu *database application* sehingga menghasilkan sebuah *conceptual database schema* pada DBMS *independent* model data tingkat tinggi seperti EER (*enhanced entity relationship*) model. Skema ini dapat dihasilkan dengan menggabungkan bermacam-macam kebutuhan *user* dan secara langsung membuat skema *database* atau dengan merancang skema-skema yang terpisah dari kebutuhan tiap-tiap *user* dan kemudian menggabungkan skema-skema tsb. Model data yang digunakan pada perancangan skema konsepsual adalah DBMS-*independent*, dan langkah selanjutnya adalah memilih sebuah DBMS untuk melaksanakan rancangan tsb.

*Entity Relationship Diagram* (ERD) digunakan untuk menggambarkan hubungan secara logika antar entitas yang terlibat pada suatu sistem *database*. Dalam kasus perancangan basisdata Sistem Informasi Penggajian Dosen, didapat minimal 7 (tujuh) buah entitas, seperti gambar berikut:

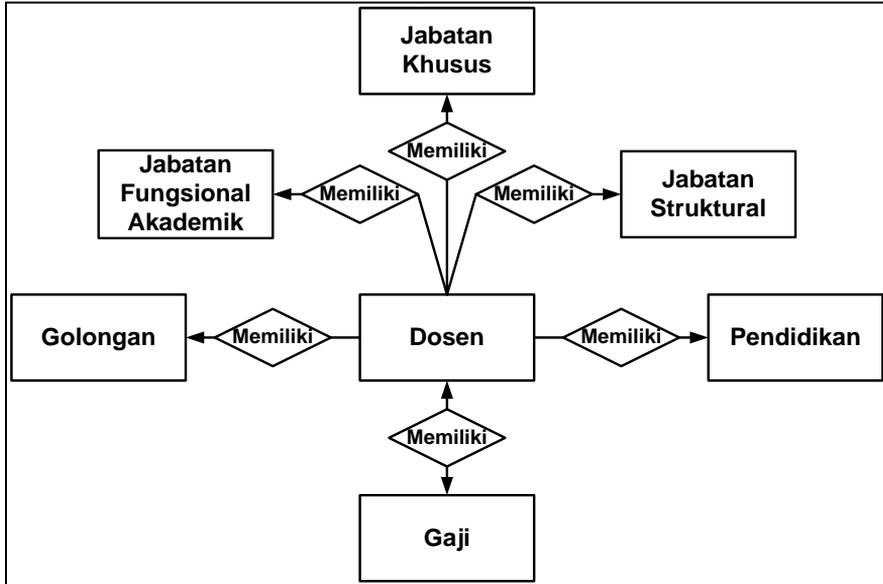

**Gambar 2. Skema Konsepsual (*Conceptual Scheme*)**

Pada skema konsepsual di atas, digunakan notasi *cardinality ratio* dengan menggunakan tanda anak panah. Sisi yang bertanda anak panah merupakan representasi dari sisi 1 (satu/*one*). Dan sebaliknya, sisi yang tidak ada tanda anak panahnya merupakan represntasikan sisi N (banyak/*many*).

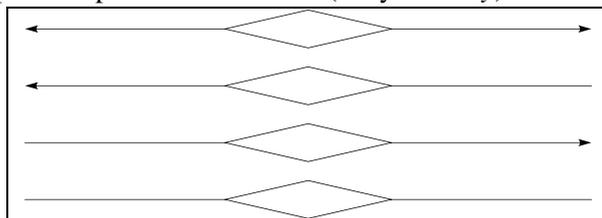

**Gambar 3. Notasi *Ratio Cardinality* (1-1, 1-N, M-1, dan M-N)**

**Perancangan Basisdata secara Logika (*Logical Design*)**

Pada fase ini, skema konsepsual ditransformasikan dari model data tingkat tinggi, untuk direpresentasikan ke dalam model data dari DBMS yang dipilih/digunakan. Untuk skema kali ini, kita gunakan *Relational Database*. Fase ini dikenal juga dengan istilah pemetaan model data (*data model mapping*).

Pada gambar 3.4, ada 7 (tujuh) tabel yang berelasi, yaitu: 1) Tabel Golongan, 2) Tabel Jabatan Fungsional Akademik, 3) Tabel Jabatan Struktural, 4) Tabel Jabatan Khusus, 5) Tabel Pendidikan, 6) Tabel Dosen, dan 7) Tabel Gaji.

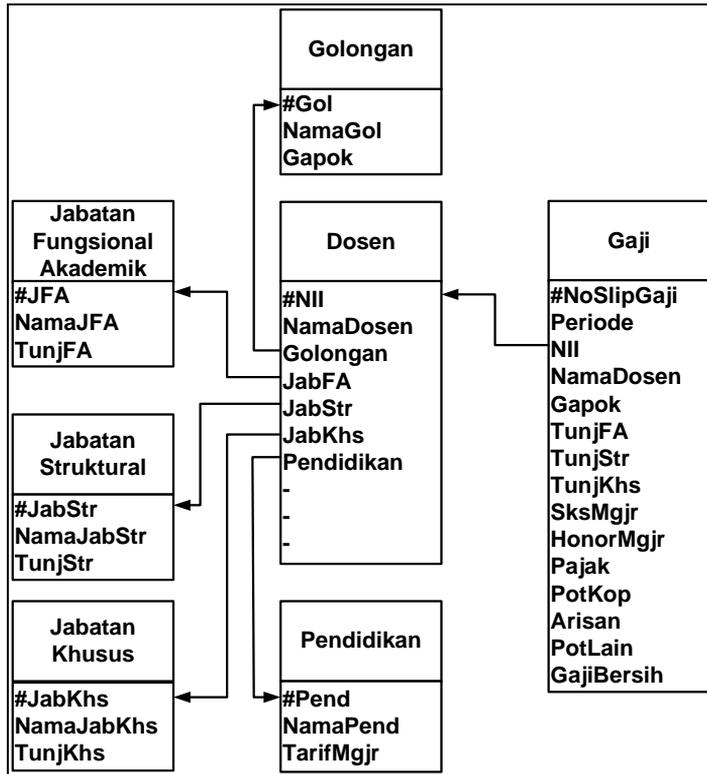

**Gambar 4. Skema Logika (*Logical Scheme*)**

**Perancangan Basisdata secara Fisik (*Physical Design*)**

Perancangan *database* secara fisik merupakan proses pemilihan struktur-struktur penyimpanan dan jalur-jalur akses pada file-file *database* untuk mencapai penampilan yang terbaik pada bermacam-macam aplikasi. Selama fase ini, dirancang spesifikasi-spesifikasi untuk database yang disimpan yang berhubungan dengan struktur-struktur penyimpanan fisik, penempatan record dan jalur akses. Berhubungan dengan internal schema (pada istilah 3 level arsitektur DBMS).

Berdasarkan Skema Logika diatas, dapat dirancang *file-file* yang digunakan sebagai penyimpanan data, masukan (*input*) yang diperlukan sistem informasi untuk menghasilkan keluaran (*output*) informasi.

Desain *File* merupakan kumpulan *record-record* yang saling berhubungan, dimana *file* tersebut dapat dimanipulasi. Dalam membuat rancangan fisik (*detail table*) dari suatu skema *Relational Database* (RD), didahulukan pada tabel yang berada pada sisi : 1) *One* (1), *One* atau *Many* (1/N), dan 3) sisi *Many* (N). Maka pada perancangan *database* untuk Sistem Informasi Penggajian ini terdapat atau menggunakan 7 (tujuh) buah *file*, yaitu 1) Golongan, 2) Jabatan Fungsional Akademik, 3) Jabatan Struktural, 4) Jabatan Khusus, 5) Pendidikan, 6) Dosen, dan 7) Gaji.

**Tabel 1. Desain Tabel Golongan**

| No. | Nama Field | Tipe Data | Lebar | Ket. |
|---|---|---|---|---|
| 1. | #Gol | Auto-Increment | - | Primary-Key |
| 2. | NamaGol | Alpha | 25 | - |
| 3. | Gapok | Currency | | - |

**Tabel 2. Desain Tabel Jabatan Fungsional Akademik**

| No. | Nama Field | Tipe Data | Lebar | Ket. |
|---|---|---|---|---|
| 1. | #JFA | Auto-Increment | - | Primary-Key |
| 2. | NamaJFA | Alpha | 30 | - |
| 3. | TunjFA | Currency | | - |

**Tabel 3. Desain Tabel Jabatan Struktural**

| No. | Nama Field | Tipe Data | Lebar | Ket. |
|---|---|---|---|---|
| 1. | #JStr | Auto-Increment | - | Primary-Key |
| 2. | NamaJStr | Alpha | 30 | - |
| 3. | TunjStr | Currency | | - |

**Tabel 4. Desain Tabel Jabatan Khusus**

| No. | Nama Field | Tipe Data | Lebar | Ket. |
|---|---|---|---|---|
| 1. | #JKhs | Auto-Increment | - | Primary-Key |
| 2. | NamaJKhs | Alpha | 30 | - |
| 3. | TunjKhs | Currency | | - |

**Tabel 5. Desain Tabel Pendidikan**

| No. | Nama Field | Tipe Data | Lebar | Keterangan |
|---|---|---|---|---|
| 1. | #Pend | AutoIncrement | - | PrimaryKey |
| 2. | NamaPend | Alpha | 30 | - |
| 3. | TarifMgjr | Currency | | - |

### Tabel 6. Desain Tabel Dosen

| No. | Nama Field | Tipe Data | Lebar | Keterangan |
|---|---|---|---|---|
| 1. | #NII | Alpha | 10 | PrimaryKey |
| 2. | NamaDosen | Alpha | 25 | - |
| 3. | Golongan | Number | - | Foreign Key |
| 4. | JabFA | Number | - | Foreign Key |
| 5. | JabStr | Number | - | Foreign Key |
| 6. | JabKhs | Number | - | Foreign Key |
| 7. | Pendidikan | Number | - | Foreign Key |
| 8. | - | - | - | - |
| 9. | - | - | - | - |
| 10. | - | - | - | - |

### Tabel 7. Desain Tabel Gaji

| No. | Nama Field | Tipe Data | Lebar | Keterangan |
|---|---|---|---|---|
| 1. | #NoSlipGaji | AutoIncrement | - | PrimaryKey |
| 2. | Periode | Alpha | 15 | |
| 3. | NII | Alpha | 10 | Foreign KEy |
| 4. | NamaDosen | Alpha | 25 | - |
| 5. | Gapok | Currency | - | - |
| 6. | TunjFA | Currency | - | - |
| 7. | TunjStr | Currency | - | - |
| 8. | TunjKhs | Currency | - | - |
| 9. | SksMgjr | Number | - | - |
| 10. | HonMgjr | Currency | - | - |
| 11. | Pajak | Currency | - | - |
| 12. | PotKop | Currency | - | - |
| 13. | Arisan | Currency | - | - |
| 14. | PotLain | Currency | - | - |
| 15. | GajiBersih | Currency | - | - |

**Alat dan Bahan**

Alat dan bahan yang digunakan pada penelitian ini terdiri atas: 1) perangkat keras (*hardware*), berupa: a) Personal komputer dengan *processor Intel Pentium IV*, b) Monitor BioStar, c) Printer *HP Deskjet 3920*, d) *Mouse,* dan e) *Keyboard.*, 2) Perangkta Lunak (*software*), berupa: a) *Microsoft Windows XP* sebagai sistem operasi, b) *Borland Delphi 7.0*, c) *Microsoft Visio*, dan d) *Paradox 7*.

**Rancangan Form**

Apabila *physical scheme* sudah dibuat, maka dengan mudah programmer dapat membuat rancangan input, kode program, serta rancangan output. Desain *input* merupakan suatu rancangan untuk memasukkan data. Berikut rancangan masing-masing form:

**Desain Input Data Golongan**

Objek/Komponen yang digunakan dalam form golongan ini adalah: 1) TForm, 2) TPanel, 3) TTimer, 4) Tiga (3) buah TLabel, 5) Tiga (3) buah TDBEdit, 6) TDBNavigator, 7) TDBGrid, dan 8) TBitBtn.

Gambar 5. Desain Input Form Golongan

**Desain Input Data Tunjangan Fungsional Akademik**

Objek/Komponen yang digunakan dalam form Tunjangan Fungsional Akademik ini adalah: 1) TForm, 2) TPanel, 3) TTimer, 4) Tiga (3) buah TLabel, 5) Tiga (3) buah TDBEdit, 6) TDBNavigator, 7) TDBGrid, dan 8) TBitBtn.

Gambar 6. Desain Input Form Tunjangan Fungsional Akademik

**Desain Input Data Tunjangan Struktural**

Objek/Komponen yang digunakan dalam form golongan ini adalah: 1) TForm, 2) TPanel, 3) TTimer, 4) Tiga (3) buah TLabel, 5) Tiga (3) buah TDBEdit, 6) TDBNavigator, 7) TDBGrid, dan 8) TBitBtn.

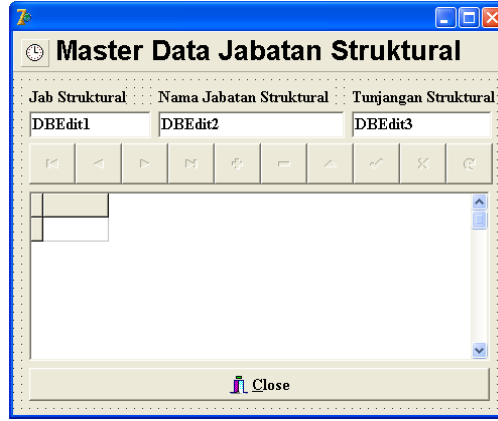

Gambar 7. Desain Input Form Jabatan Struktural

**Desain Input Data Tunjangan Khusus**

Objek/Komponen yang digunakan dalam form golongan ini adalah: 1) TForm, 2) TPanel, 3) TTimer, 4) Tiga (3) buah TLabel, 5) Tiga (3) buah TDBEdit, 6) TDBNavigator, 7) TDBGrid, dan 8) TBitBtn.

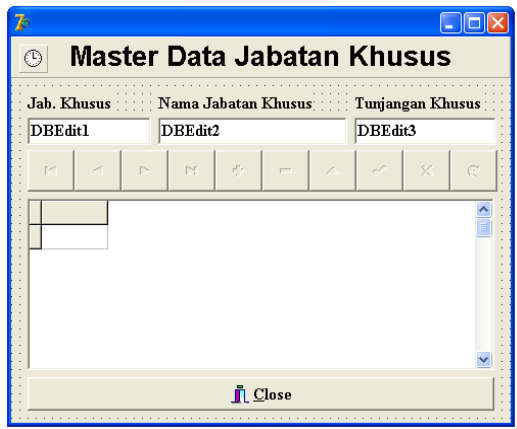

Gambar 8. Desain Input Tunjangan Khusus

**Desain Input Data Pendidikan**

Objek/Komponen yang digunakan dalam form golongan ini adalah: 1) TForm, 2) TPanel, 3) TTimer, 4) Tiga (3) buah TLabel, 5) Tiga (3) buah TDBEdit, 6) TDBNavigator, 7) TDBGrid, dan 8) TBitBtn.

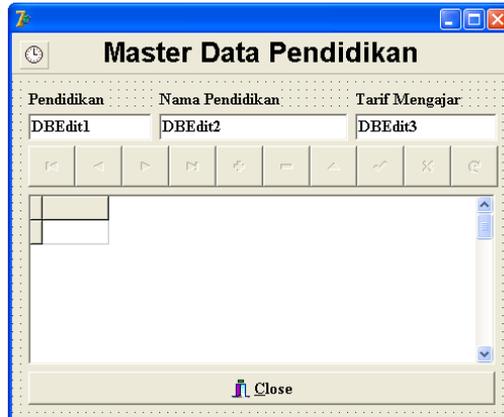

Gambar 9. Desain Input Form Pendidikan

**Desain Input Data Dosen**

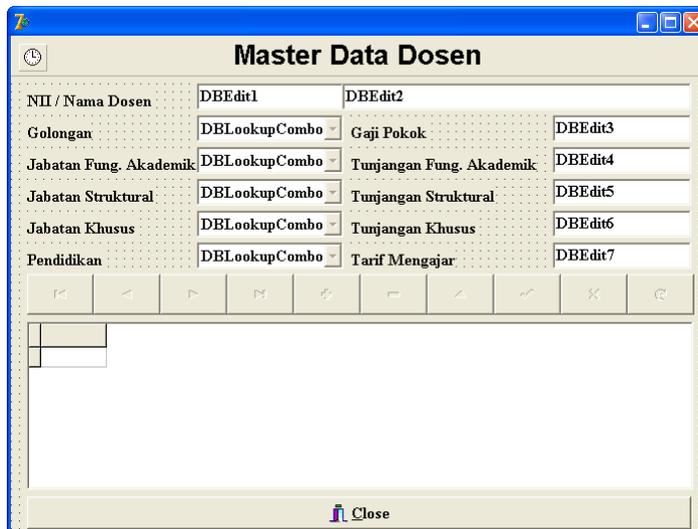

Gambar 10. Desain Input Data Dosen

Objek/Komponen yang digunakan dalam form golongan ini adalah: 1) TForm, 2) TPanel, 3) TTimer, 4) Sebelas (11) buah TLabel, 5) Tujuh (3) buah TDBEdit, 6) TDBNavigator, 7) TDBGrid, 8) TBitBtn, dan 9) Lima (5) TDBLookUpComboBox.

**Desain Input Gaji Dosen**

Objek/Komponen yang digunakan dalam form golongan ini adalah : 1) TForm, 2) TPanel, 3) TTimer, 4) Delapan belas (18) buah TLabel, 5) Tiga belas (13) buah TDBEdit, 6) TDBNavigator, 7) TDBGrid, 8) TBitBtn, 9) Lima (5) TDBLookUpComboBox, 10) Satu (1) buah TEdit, 11) Satu (10) TDateTimePicker, dan 12) Dua (2) buah TGroupBox.

**Gambar 11. Desain Input Form Dosen**

**Pembuatan dan Pengujian Program**

Sistem Informasi Penggajian ini dibuat dengan menggunakan program Borland Delphi, proses pembuatan programan adalah dengan menerjemahkan rancangan yang ada menjadi form dan report dengan objek/komponen yang disediakan oleh Borlnad Delphi.

Adapun tahapan pembuatan program dengan menggunakan Borland Delphi: 1) **Rancang Form** sesuai dengan desain, 2) **Pengaturan Setting Properties** dari masing-masing objek/komponen yang digunakan dalam form, 3) Masukkan kode program atau **Procedures** ke dalam form sesuai dengan event/aksi yang digunakan dalam form. Abdillah (2005b:1).

## 4. PEMBAHASAN

Setelah melakukan pembuatan program, penelitian ini menghasilkan 1 (sebuah) project, 1 (sebuah) *database*, 6 (form) form, dan 5 (lima) report. Berikut akan penjelasannya.

### Form Menu Utama dan Sub Menu

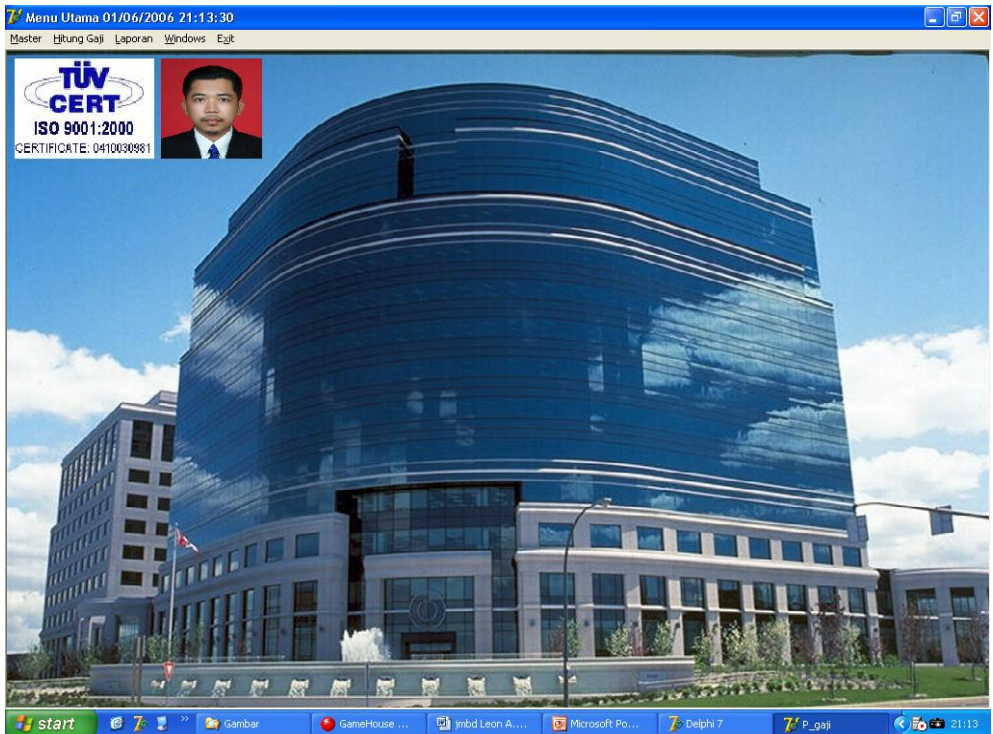

**Gambar 12. Main Menu dan Sub Menu**

**Form Data Dosen**

Gambar 13. Form Data Dosen

**Form Gaji**

Gambar 4.3 Form Gaji

5. **KESIMPULAN**

Berdasarkan hasil penelitian ini dapat diambil beberapa kesimpulan tentang sisteminformasi inventarisasi barang pada Universitas Bina Darma Palemabng sebagai berikut:
1) Dengan perancangan ini, diharapkan dapat dikembangkan menjadi suatu program utuh untuk mengolah data gaji dengan akurat, cepat, efisien.
2) Skema yang dihasilkan dapat diadopsi ke sejumlah sistem penggajian dosen pada perguruan tinggi lainnya.
3) Perancangan Basisdata Sistem Informasi Penggajian, menghasilkan 7 (tujuh) buah tabel, 6 (enam) buah form, dibuat dengan menggunakan salah satu bahasa pemrograman berorientasi *object* yaitu *Borland Delphi* 6.0. menggunakan aplikasi *database* dengan format *Paradox*.

# DAFTAR RUJUKAN

Abdillah, Leon, Andretti, 2003, *Sistem Basis Data Lanjut 1: Membangun Sistem Basis Data*, Universitas Bina Darma, Palembang.

Abdilllah, Leon, Andretti, 2004, Sistem Informasi Inventaris Barang, *Jurnal Ilmiah MATRIK*, 6(3):133-152, Palembang

Abdillah, Leon, Andretti, 2005a, *Pemrograman III (Delphi Database)*, Universitas Bina Darma, Palembang.

Abdillah, Leon, Andretti, 2005b, Validasi Data dengan menggunakan Objek Lookup pada Borland Delphi 7.0, *Jurnal Ilmiah MATRIK*, 7(1):1-16, Palembang.

Abdillah, Leon, Andretti, 2006, *Pengaruh Kompensasi dan Teknologi Informasi terhadap Kinerja Dosen (KIDO) Tetap*, Tesis Tidak Dipublikasikan, Program Pascasarjana Magister Manajemen, Universitas Bina Darma, Palembang.

Cornelia J. Benny, C. J., 2004. *Pengembangan Manajemen Dosen di Perguruan Tinggi Sebagai Upaya Meningkatkan Mutu Kinerja Dosen, Disertasi Dipublikasikan*, (OnLine), (www.pages-yourfavorite.com/ppsupi/ diakses pada Desember 2005).

Indrajit, R. E. 20001. *Manajemen Sistem Informasi dan Teknologi Informasi*, Jakarta: Elex Media Komputindo, Jakarta.

Jogianto, H. 2000. *Analisis dan Desian Sistem Informasi: Pendekatan Terstruktur Teori dan Praktek Aplikasi Bisnis*, Andi, Yogyakarta.

Koordinator Kopertis Wilayah II, 2005, *Pedomam Umum Pengusulan Kenaikan Pangkat dan Jabatan Fungsional Dosen Perguruan Tinggi Swasta Kopertis Wilayah II*, Palembang: Departemen Pendidikan Nasional Koordinasi Perguruan Tinggi Swasta Wilayah II, Palembang.

Nur, E., 2001, *Sistem Penggajian*, (OnLine), (http://kuliah.dinus.ac.id/edi-nur/lembar01.html/ lembar01.html diakses pada Desember 2005).

Sekarwati, K. A. 2001. *Sistem Basis Data*. (OnLine). (http://staffsite.gunadarma.ac.id/, diakses 28 Februari 2003).

Wardiana, W. 2004. *Perkembangan Teknologi Informasi di Indonesia*, (OnLine), (http://eprints.rclis.org/archive/ diakses pada Desember 2005).